\documentclass[final,english]{bullsrsl}[2022/06/15]

\usepackage[latin1]{inputenc}
\usepackage[T1]{fontenc}
\usepackage{natbib} 
\usepackage{graphicx}

\begin{document}
\title{Evolution and Final Fates of a Rotating 25\,M$_{\odot}$ Pop III star}

\author[affil={1,2}, corresponding]{Amar}{Aryan}
\author[affil={1}]{Shashi Bhushan}{Pandey}
\author[affil={1,2}]{Rahul}{Gupta}
\author[affil={2}]{Sugriva Nath}{Tiwari}
\author[affil={1}]{Amit Kumar}{Ror}
\affiliation[1]{Aryabhatta research institute of observational sciences (ARIES), Nainital, Uttarakhand, India-263001}
\affiliation[2]{Department of Physics, Deen Dayal Upadhyaya Gorakhpur University, Gorakhpur, Uttar Pradesh, India-273009}
%\affiliation[3]{Institute of Improbability, Impossible City}
%\affiliation[4]{Double Blind Testing Inc., Great Opportunities}
\correspondance{amararyan941@gmail.com and amar@aries.res.in}
\date{31st May 2023}

\maketitle

% \author[affil1]{FirstName (+ MiddleInitials if necessary)}{FamilyName}
% \author[affil2]{...}{}
% \equalcontribauthor[]{}{} % Maximum two --> counter
% \consortium[affil]{Consortium Name}
% With consortium: affiliation will be set to "See Appendix 1 for a full
% list of consortium members and their respective affiliations
% \affiliation[affil1]{...}
% \affiliationq[affil2]{...}

% \correspondence[]{}
% No explicit corresponding author: use first author
% 

% Abstract of the paper in the same language as the paper
\begin{abstract}
In this proceeding, we present the 1-dimensional stellar evolution of two rotating population III (Pop III) star models, each having a mass of 25\,M$_{\odot}$ at the zero-age main-sequence (ZAMS). The slowly rotating model has an initial angular rotational velocity of 10 per cent of the critical angular rotational velocity. In contrast, the rapidly rotating model has an initial angular rotational velocity of 70 per cent of the critical angular rotational velocity. As an effect of rotationally enhanced mixing, we find that the rapidly rotating model suffers an enormous mass loss due to the deposition of a significant amount of CNO elements toward the surface after the main-sequence phase. We also display the simulated light curves as these models explode into core-collapse supernovae (CCSNe).
\end{abstract}

\keywords{Population III stars, Stellar evolution, Hydrodynamic simulation, Supernovae}

%\section{Section -- Level 1 title (Times New Roman, bold, 14 pts)}
\section{Introduction}
\label{Intro}
Pop III stars refer to the first generation of stars, a captivating and enigmatic class of astrophysical objects that were thought to be born in the early Universe before the formation of any other stars. These primordial stars are believed to have formed from initial, pristine gas composed almost entirely of Hydrogen and Helium, lacking any heavier elements {\bf \citep[][]{1981ApJ...248..606B,1986A&A...168...81C}}. Because of their unique composition and lack of any coolant in the early Universe, Pop III stars are thought to have been much more massive than stars in the later generations \citep[][]{2015MNRAS.448..568H}. They played a crucial role in shaping the Universe as their intense radiation ionized the surrounding gas to initiate the process of cosmic reionization \citep[][]{2013RPPh...76k2901B}. While no Population III stars have been directly observed yet, their existence is supported by theoretical models \citep[e.g., among many others][]{1999ApJ...515..239N, 2007ApJ...654...66O} and indirect evidence \citep[e.g., among many others][]{2014ApJ...792...44C,2015MNRAS.450.2506V,2016MNRAS.462..601R,2016ApJ...823...83M,2018MNRAS.478.5591M}. The studies related to these first-ever stellar objects are the key to unveil the mysteries of the early Universe and are also very important to understand the origins of other Pop II and Pop I stars. There are multiple studies to understand the possible existence, evolution, and final fates of Pop III stars \citep[e.g., among many others,][]{2003A&A...399..617M,2008A&A...489..685E,2009Sci...325..601T,2012A&A...542A.113Y,2014ApJ...781...60H,2016ApJ...826....9I,2018ApJS..234...41W,2021MNRAS.501.2745M,2023MNRAS.521L..17A}. In this work, we investigate the cause of enormous mass loss in rapidly rotating model as it passes through various stages of its evolution. We also present the hydrodynamic simulations of synthetic explosions of the models at the onset of core collapse.

We have divided this proceeding into four sections. We present a brief overview of the literature in Section~\ref{Intro}. The numerical settings of the models to perform their stellar evolution are presented in Section~\ref{Modelling} while the methods to simulate the synthetic explosions are discussed in Section~\ref{Explosions}. Finally, we present our results and conclusions in Section~\ref{Results}.
\begin{figure}
\centering
    \includegraphics[height=6.7cm,width=0.46\columnwidth,angle=0]{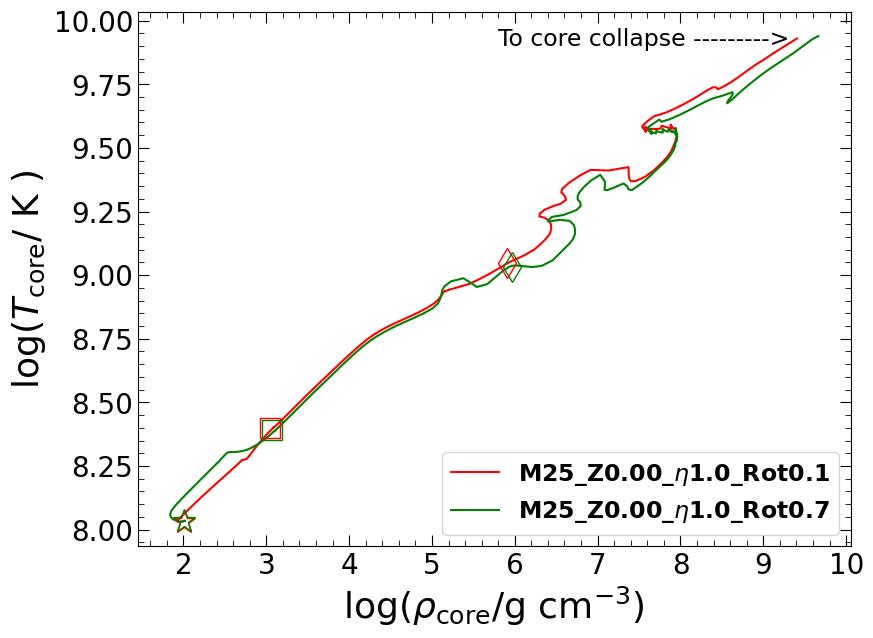}
    \includegraphics[height=7.3cm,width=0.46\columnwidth,angle=0]{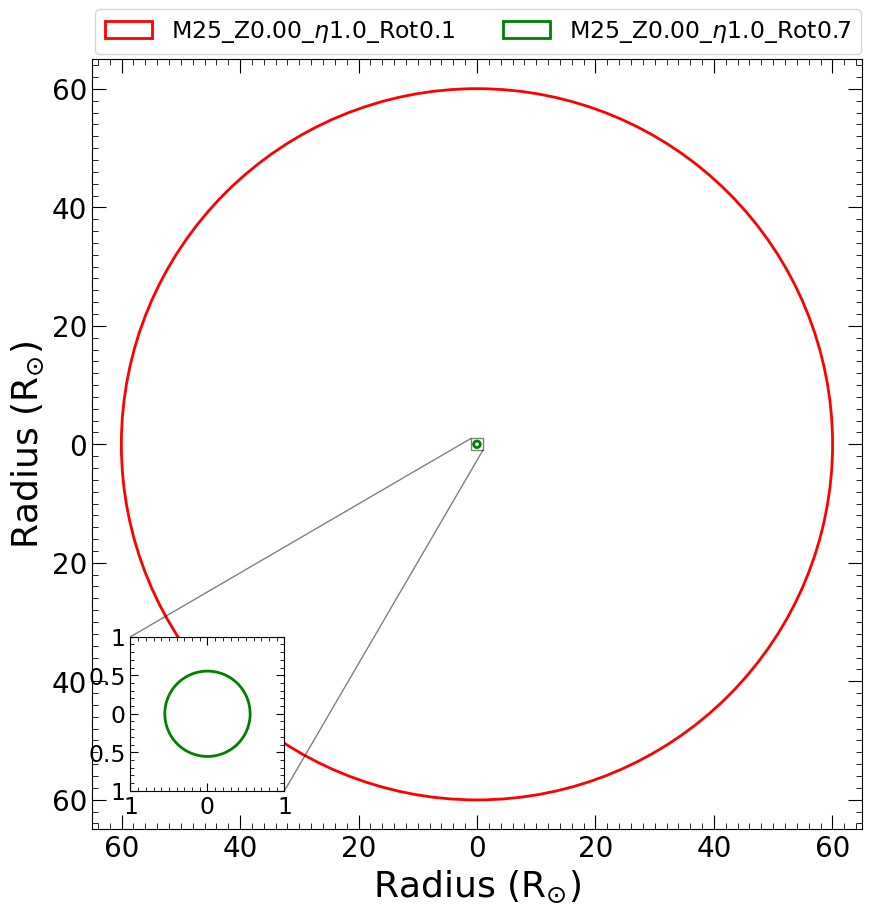}
   \caption {{\em Left:} The evolution of the $T_{\rm core}$ vs $\rho_{\rm core}$ curves as our Pop III models evolve on the HR diagram. The phases of arrival on ZAMS, the exhaustion of core-He burning, and the exhaustion of core-C burning are indicated by hollow stars, squares, and diamonds, respectively. {\em Right:} The Pre-SN radii of the models. The rapidly rotating model seems to have suffered enormous mass loss and thus possesses a very small Pre-SN radius.}
    \label{fig:Rho_Radius}
\end{figure}

\begin{figure}
\centering
    \includegraphics[height=5.0cm,width=0.49\columnwidth,angle=0]{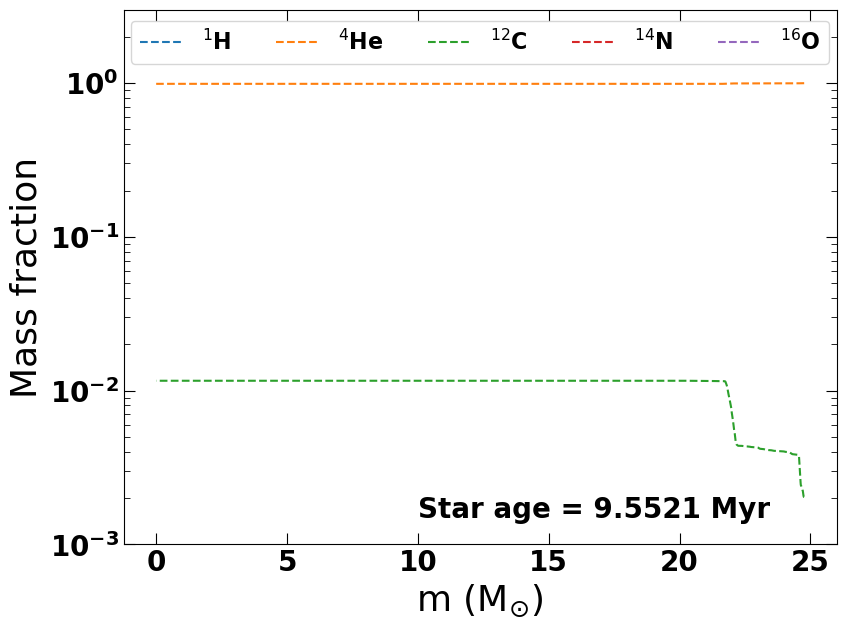}
    \includegraphics[height=5.0cm,width=0.49\columnwidth,angle=0]{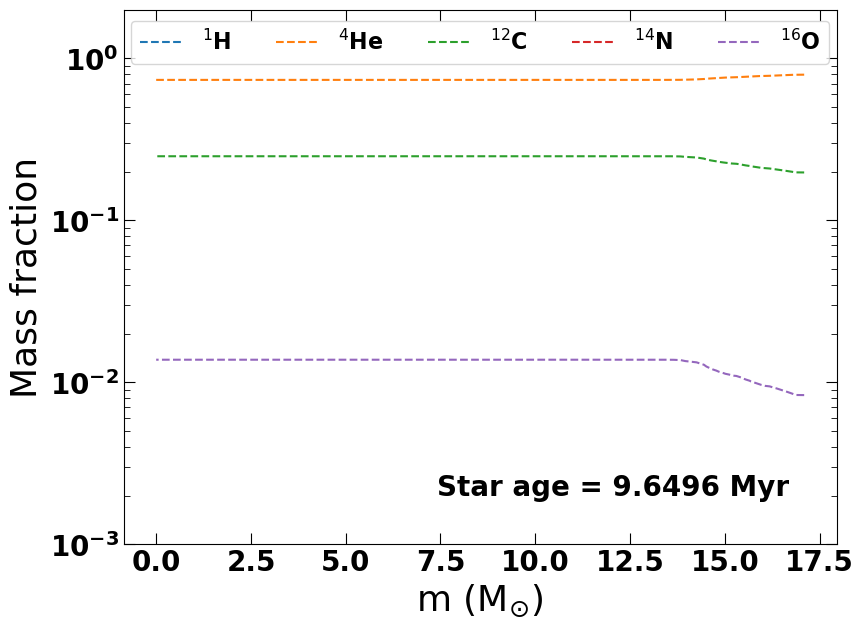}
    \includegraphics[height=5.0cm,width=0.49\columnwidth,angle=0]{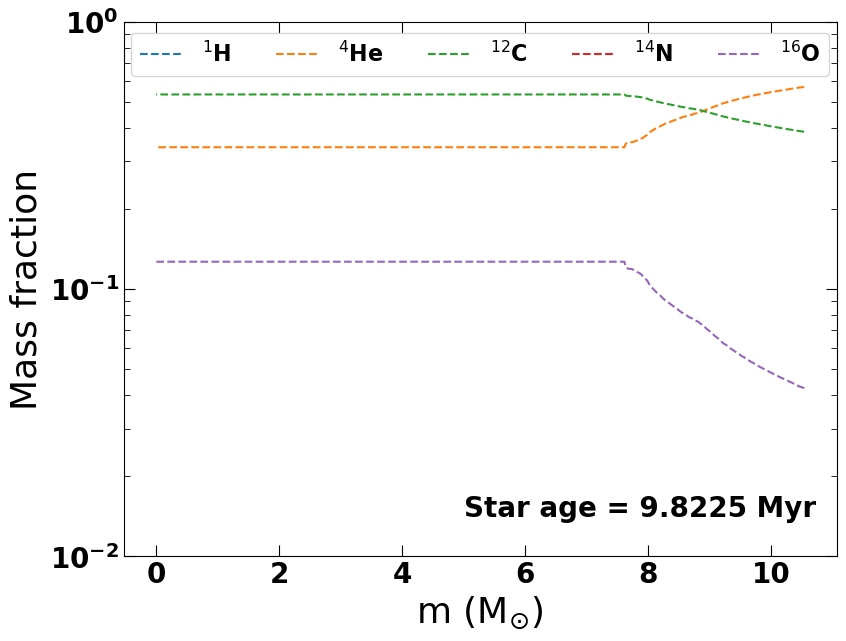}
    \includegraphics[height=5.0cm,width=0.49\columnwidth,angle=0]{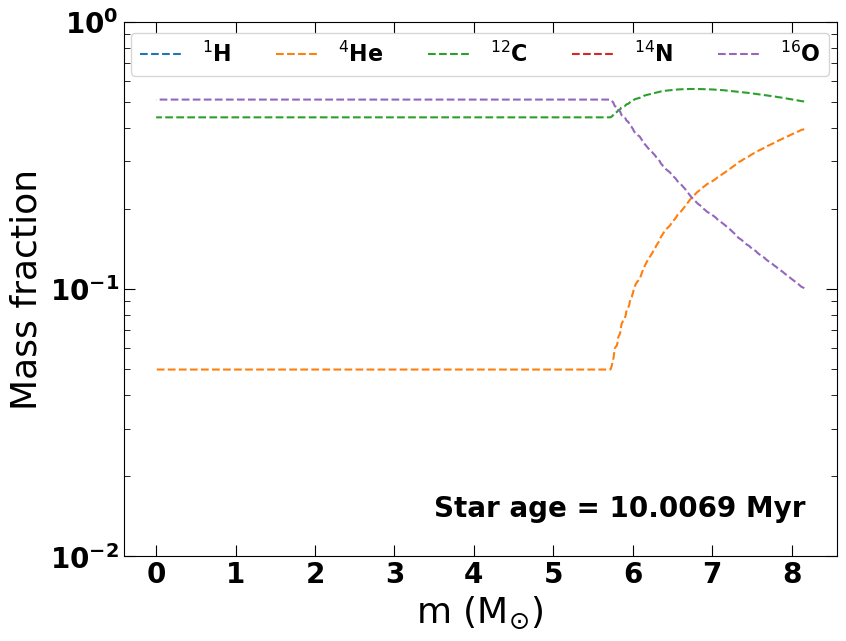}    
   \caption {This set of four figures shows the mass fraction of the rapidly rotating model (M25\_Z0.00\_$\eta$1.0\_Rot0.7) at four epochs after the main-sequence phase. The increased fractions of CNO elements near the surface dramatically boost the surface metallicity.}
    \label{fig:Mass_frac}
\end{figure}

\section{Evolution of the models upto Pre-SN stage}
\label{Modelling}
To perform the stellar evolution of the models, we utilise the modules for experiments in stellar astrophysics ({\tt MESA}) with version number mesa-r21.12.1 \citep[][]{2011ApJS..192....3P,2013ApJS..208....4P,2015ApJS..220...15P,2018ApJS..234...34P}. Under the present work, we take two 25\,M$_{\odot}$ ZAMS star models with zero metallicity and perform their 1-dimensional stellar evolution until they reach the onset of the core-collapse phase. The models are marked to have reached the onset of the core-collapse stage if any location within the star model hits an infall velocity of 500\,km\,s$^{-1}$. The {\tt MESA} settings for the calculations presented here are similar to the ones used in \citet[][]{2021RMxAC..53..215A,2022JApA...43...87A,2022JApA...43....2A} and closely follow \citet[][]{2023MNRAS.521L..17A}. However, we list a few critical changes. We have performed the stellar evolution of two models with initial rotations ($\Omega/\Omega_{\rm crit}$) of 0.1 and 0.7, respectively. In this work, we have also investigated the effect of changing the wind scaling factor ($\eta$) from 0.5 to 1.0. The models presented in this work are so named that they contain information on initial ZMAS mass, metallicity, scaling factor, and rotation. The slowly rotating model named M25\_Z0.00\_$\eta$1.0\_Rot0.1 indicates a star with ZAMS mass of 25\,M$_{\odot}$, zero metallicity, $\eta$ equals to 1.0, and an initial rotation of 0.1. Similarly, the rapidly rotating model named M25\_Z0.00\_$\eta$1.0\_Rot0.7 indicates a star with ZAMS mass of 25\,M$_{\odot}$, zero metallicity, $\eta$ equals to 1.0, and an initial rotation of 0.7. The left-hand panel of Figure~\ref{fig:Rho_Radius} shows the variation of the core-temperature ($T_{\rm core}$) vs core-density ($\rho_{\rm core}$) curve as the models evolve from the ZAMS to core-collapse phase. The Pre-SN parameters are mentioned in Table~\ref{tab:MESA_MODELS}. The right-hand panel of Figure~\ref{fig:Rho_Radius} shows the Pre-SN radii of the two models. The rapidly rotating model has undergone significant mass loss, resulting in a very small Pre-SN radius. In contrast, the slowly rotating model has retained most of its outer Hydrogen-envelope. Another effect evident as a result of increasing the $\eta$ from 0.5 to 1.0 is an increased amount of lost mass in the rapidly rotating model considered here. Although the M25\_Z0.00\_Rot0.8 model from \citet[][]{2023MNRAS.521L..17A} has a more initial rotation than the M25\_Z0.00\_$\eta$1.0\_Rot0.7 model here, the later has lost much more mass than earlier due to an increased $\eta$.
Additionally, we find that the rapidly rotating model has suffered an enormous mass loss compared to the slowly rotating model. We have performed a diagnosis to explain this enormous mass loss. The four panels of Figure~\ref{fig:Mass_frac} display the mass fractions of several elements after the main-sequence phase. As the model progresses on the HR diagram beyond the main-sequence, the fractions of CNO elements toward the surface increase, dramatically enhancing surface metallicity. The increased surface metallicity, in turn, enhances mass loss \citep[][]{2007A&A...461..571H}.     

\begin{figure}
\centering
    \includegraphics[height=6.5cm,width=0.49\columnwidth,angle=0]{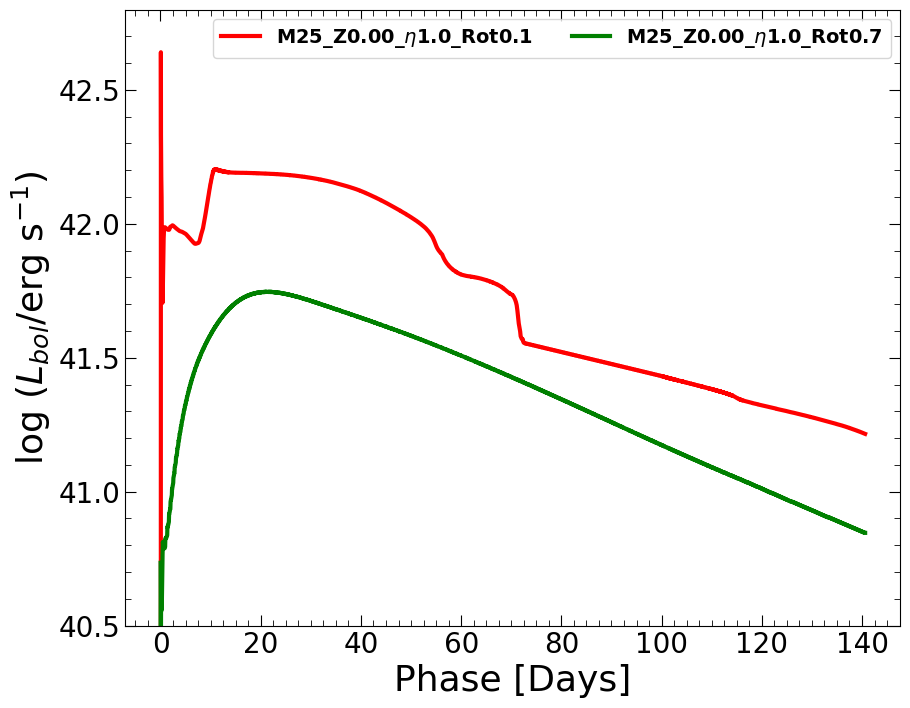}
    \includegraphics[height=6.5cm,width=0.49\columnwidth,angle=0]{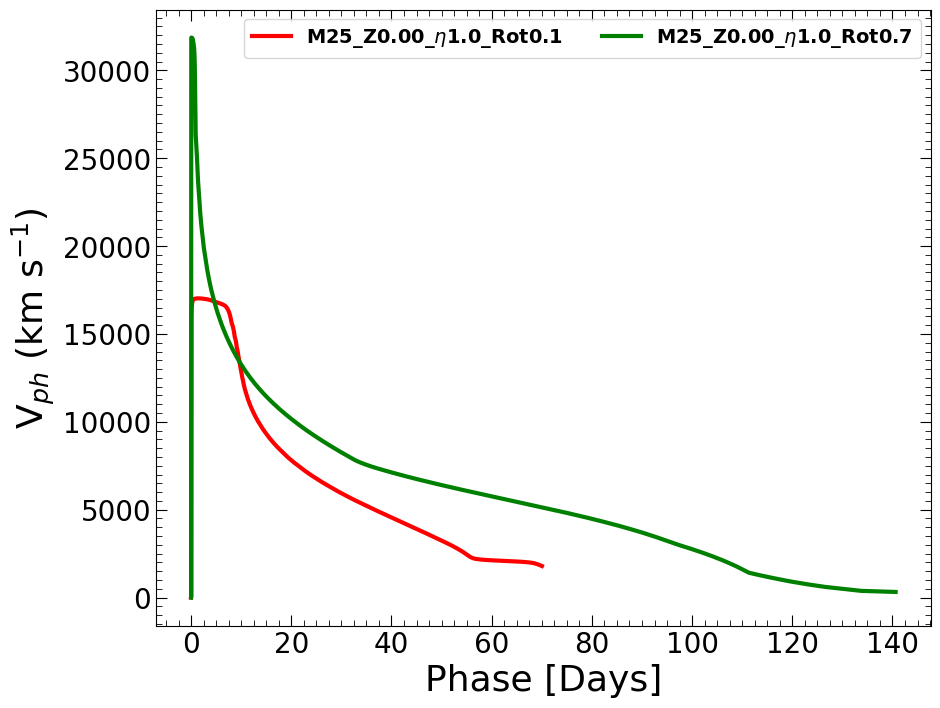}
   \caption {{\em Left:} The bolometric luminosity light curves of the two models considered in this work, resulting from the hydrodynamic simulations of their synthetic explosions utilising {\tt SNEC}. {\em Right:} The evolution of corresponding photospheric velocities obtained from {\tt SNEC}.}
    \label{fig:Lum_Vel}
\end{figure}

\section{Explosion of the Pre-SN models}
\label{Explosions}
Once the models reach the core-collapse stage, we simulate their synthetic explosions utilising {\tt SNEC} \citep[][]{2015ApJ...814...63M}. Most of the {\tt SNEC} settings are similar to those in \citet[][]{2019ApJ...877...92O,2021MNRAS.505.2530A,2022MNRAS.517.1750A,2023MNRAS.521L..17A}. Here, we mention important modifications. We choose the ``Piston\_Explosion" option to simulate the synthetic explosion with a set of 700 grid cells using {\tt SNEC}. {\bf  For CCSNe, the ``Piston\_Explosion" might be the more realistic one since these SNe are thought to be arising due to the shock wave bouncing back from the neutron star. On the other hand, the ``Thermal\_Bomb" type is better suited for thermonuclear explosions, like Type Ia SNe.} As we utilise the "Piston\_Explosion" in SNEC, the first two computational cells in our model's profile are subjected to an outward velocity boost (in cm\,s$^{-1}$) provided by the "piston\_vel" control. We choose "piston\_vel = 4d9" for both models. The period of velocity boost lasts for 0.01\,s. 
\begin{table*}
\caption{The Pre-SN properties of the two models using {\tt MESA}. The corresponding {\tt SNEC} explosion parameters are also presented here. }
\label{tab:MESA_MODELS}
\begin{center}
{\scriptsize
\begin{tabular}{ccccccccccccc}
\hline \hline

  & & Pre-SN  & & \hspace{2.3cm}\vline & &  Explosion\\
\hline
Model Name	&	$M_{\rm Pre-SN}^{a}$	& $T_{\mathrm{eff}}$  & $R_{\mathrm{Pre-SN}}^{b}$  	& log$_{10}$($L_{\rm Pre-SN}$/L$_{\odot}$)  &	$M_{\mathrm{c}}^{d}$ & $M_{\mathrm{ej}}^{e}$  &	$V_{\mathrm{boost}}^{f}$ & $E_{\mathrm{exp}}^{g}$\\
	&     (M$_{\odot}$) & K	&	(R$_{\odot}$)&	 & (M$_{\odot}$) & (M$_{\odot}$)  &	($10^{9}$\,cm\,s$^{-1}$) & ($10^{51}$\,erg)\\ 	
\hline
\hline

M25\_Z0.00\_$\eta$1.0\_Rot0.1     &	 24.99  &    16172  & 60.04  & 5.345 &  1.60 & 23.39  & 4.0	& 6.58\\

M25\_Z0.00\_$\eta$1.0\_Rot0.7     & 7.71  &    167875  & 0.55 & 5.340 & 1.50 & 6.21  & 4.0	& 4.07	\\
		
\hline\hline
\end{tabular}}
\end{center}
%\par
{
$^a$Mass at the Pre-SN stage,
$^b$Radius at the Pre-SN stage,
$^c$Luminosity at the Pre-SN stage,
$^d$Excised central remnant mass,
$^e$Ejecta mass,
$^f$Boosting-velocity of the first two computational cells in the model profile,
$^g$Explosion energy.}\\

\end{table*} 
For each model considered in this work, we first excise the mass of the final remnant ($M_{\rm c}$), which is nearly the mass of the inert Iron-core. Additionally, we use an amount of 0.05\,M$_{\odot}$ of $^{56}$Ni synthesised for both the models. This quantity of synthesised $^{56}$Ni is distributed between the excised central remnant mass cut and the preferred mass coordinate, which is in close proximity to the outer surface of the models. The difference in the Pre-SN mass ($M_{\rm Pre-SN}$) and the $M_{\rm c}$ is the corresponding ejecta mass for each model. We present the detailed explosion parameters in Table~\ref{tab:MESA_MODELS}. The left panel of Figure~\ref{fig:Lum_Vel} shows the bolometric luminosity light curves for the two models. The slowly rotating model has retained most of its outer Hydrogen-envelope; thus, its explosion results in a Hydrogen-rich SN. The bolometric light curve closely resembles the Type IIP SNe light curves. In contrast, the rapidly rotating model has suffered extensive mass loss. Thus, it explodes as a Hydrogen-stripped SN. The bolometric light curve from the rapidly rotating model mimics the light curves of Hydrogen-deficient Type Ib/c SNe. The right-hand panel of Figure~\ref{fig:Lum_Vel} displays the corresponding photospheric velocity evolution for the two models. The slowly rotating model resembles the photospheric velocities shown by Type IIP SNe, while the very high initial photospheric velocities resemble stripped-envelope SNe.  

\section{Results and Conclusions}

In this proceeding, we performed the 1-dimensional stellar evolution of two rotating Pop III models until they reached the stage of the onset of core collapse utilising {\tt MESA}. Further, we also performed the hydrodynamic simulations of their synthetic explosions using the models at the onset of core collapse in appropriate form as input to {\tt SNEC}. We enumerate our findings below:
\begin{enumerate}
\item We explicitly explored the cause of extensive mass loss in our rapidly rotating model by investigating the mass fraction plots at different stages after the main-sequence phase. We find that the increase in mass loss rates can be attributed to the dramatic increase in surface metallicity.

\item We found that increasing $\eta$ from 0.5 to 1.0 also played an essential role in increasing the mass loss.  

\item Unlike \citet[][]{2023MNRAS.521L..17A}, in this work, we simulated the piston-driven explosion. However, we hardly see much difference in our results.
\end{enumerate}
\label{Results}

\begin{acknowledgments}
{\bf We thank the anonymous referee for providing constructive comments. We acknowledge the Belgo-Indian Network for Astronomy and astrophysics (BINA) consortium approved by the International Division, Department of Science and Technology (DST, Govt. of India; DST/INT/BELG/P-09/2017) and the Belgian Federal Science Policy Office (BELSPO, Govt. of Belgium; BL/33/IN12), for allowing us to present our work in the form of a proceeding.} We duly acknowledge the extensive utilisation of ARIES's High-Performance Computing (HPC) facility. A.A. duly acknowledges the funds and support provided by the Council of Scientific \& Industrial Research (CSIR), India, under file no. 09/948(0003)/2020-EMR-I. SBP and RG duly acknowledge the funds and support furnished by the Indian Space Research Organisation (ISRO) under the AstroSat archival Data utilisation grant DS$\_$2B-13013(2)/1/2021-Sec.2.
\end{acknowledgments}

\begin{furtherinformation}

\begin{orcids}
\orcid{0000-0002-9928-0369}{Amar}{Aryan}
%\orcid{1111-2222-3333-4444}{Shashi Bhushan}{Pandey}
\orcid{0000-0003-4905-7801}{Rahul}{Gupta}
\end{orcids}

\end{furtherinformation}

\bibliographystyle{bullsrsl-en}
\bibliography{extra}

\end{document}